\documentclass[sigconf]{acmart}
\usepackage{xspace}
\usepackage{algorithm}
\usepackage{algpseudocode}
\usepackage{marvosym}
\AtBeginDocument{%
  }

\setcopyright{acmlicensed}
\copyrightyear{2018}
\acmYear{2018}
\acmDOI{XXXXXXX.XXXXXXX}
\acmConference[Conference acronym 'XX]{Make sure to enter the correct
  conference title from your rights confirmation email}{June 03--05,
  2018}{Woodstock, NY}
\acmISBN{978-1-4503-XXXX-X/2018/06}

\newcommand{\model}{HubMixer\xspace}
\begin{document}

\title{\model: Progressive Latent Hub Mixing for Parameter-Efficient Feature Interaction in Recommendation}



\author{Jie Zhou}
\affiliation{
  \institution{Kuaishou Technology}
  \country{Beijing, China}
}
\email{
  zhoujie15@kuaishou.com
}

\author{Zixian Gong}
\affiliation{
  \institution{Tsinghua University}
  \country{Beijing, China}
}
\email{
  gong-zx22@mails.tsinghua.edu.cn
}

\author{Wenhao Li}
\affiliation{
  \institution{Kuaishou Technology}
  \country{Beijing, China}
}
\email{
  liwenhao05@kuaishou.com
}

\author{Chang Liu}
\affiliation{
  \institution{Kuaishou Technology}
  \country{Beijing, China}
}
\email{
  liuchang41@kuaishou.com
}

\author{Enzhao Shen}
\affiliation{
  \institution{Kuaishou Technology}
  \country{Beijing, China}
}
\email{
  shenenzhao@kuaishou.com
}

\author{Bo Liu}
\affiliation{
  \institution{Kuaishou Technology}
  \country{Beijing, China}
}
\email{
  liubo22@kuaishou.com
}

\author{Xu Guo\textsuperscript{\Letter}}
\affiliation{
  \institution{Kuaishou Technology}
  \country{Beijing, China}
}
\email{
  guoxu08@kuaishou.com
}


\author{Fei Pan}
\affiliation{
  \institution{Kuaishou Technology}
  \country{Beijing, China}
}
\email{
  panfei05@kuaishou.com
}

\author{Peng Jiang}
\affiliation{
  \institution{Kuaishou Technology}
  \country{Beijing, China}
}
\email{
  jiangpeng@kuaishou.com
}
\renewcommand{\shortauthors}{Jie Zhou et al.}

\begin{abstract}
    Learning effective feature interactions is central to industrial recommendation and advertising ranking systems. Recent token-mixing architectures simplify Transformer-style self-attention with lightweight mixing operators, improving hardware efficiency and enabling large-scale deployment. However, recommendation tokens are fundamentally heterogeneous: user profiles, item attributes, behavioral sequences, context features, statistical signals, and business-side features live in different semantic spaces and interact in sparse, sample-specific patterns. Directly mixing all tokens in the raw heterogeneous token space may therefore be parameter-inefficient, as the model must implicitly discover which feature groups should interact and how such interactions should be routed.

In this paper, we propose \model, a parameter-efficient latent hub mixing architecture for feature interaction in recommendation. Instead of directly mixing raw feature tokens, \model introduces a small set of learnable latent hubs to organize feature interactions through an \emph{induction--interaction--readout} paradigm. First, hub induction summarizes heterogeneous tokens into compact latent hubs, where latent hubs query input tokens through cross-attention. Second, hub interaction performs high-order interaction in the cleaner latent hub space. Third, token-conditioned readout lets each original token selectively read from the interacted hubs, injecting global interaction semantics while preserving token-level field identity. The readout signal is injected back into token representations through a residual connection, enabling stacked blocks to progressively refine cross-feature semantics. Extensive offline experiments on industrial recommendation tasks show that \model outperforms strong feature interaction and token-mixing baselines with fewer parameters. Ablation studies validate the effectiveness of hub interaction, and token-conditioned readout. Online A/B testing in the Kuaishou short-video recruitment business further shows a statistically significant 5.48\% improvement in resume submission conversion rate, and \model has been fully deployed in production.
\end{abstract}


\maketitle

\section{Introduction}
Industrial recommender systems achieve accurate prediction by modeling feature interactions over large-scale sparse and dense features. In modern recommendation ranking models, the input features typically include user profiles, user behavior histories, item features, contextual features, and real-time feedback signals. These features are commonly transformed into dense embeddings and represented as a set of tokens, where token interaction modeling directly affects the final ranking performance.

A recent line of work \cite{zhu2025rankmixer, jiang2026tokenmixer} explores Transformer-like architectures for recommendation, such as self-attentive feature interaction models and token-mixing ranking backbones. Standard self-attention provides content-adaptive token-token interaction, but its quadratic cost can be prohibitive when the token set grows. To improve efficiency, token-mixing architectures replace attention with lightweight mixing operators or matrix-friendly transformations. These mixing models are attractive for industrial deployment because they reduce computational overhead, improve hardware utilization, and scale well as the model size increases.

However, recommendation tokens are substantially different from homogeneous language tokens in NLP. Natural language tokens usually come from a shared semantic space and are arranged in a meaningful sequence. In contrast, recommendation feature tokens are highly heterogeneous: a user-age token, a historical-click sequence token, an item-category token, a context token, and a statistical token carry different semantics, have different sparsity patterns, and participate in different types of interactions. Moreover, useful interactions are usually sparse and sample-dependent. For example, user behavior tokens are often informative only when matched with item-category or job-intent tokens, while context tokens may mainly modulate time-sensitive or scenario-specific features. Directly mixing all tokens in the raw heterogeneous token space makes it challenging for the model to uncover useful semantic interactions.

This observation motivates a different question: instead of directly mixing heterogeneous tokens, can we first organize them into a small number of latent semantic hubs, perform high-order interaction in the hub space, and then inject the interacted semantics back into token representations? Such a mechanism could provide a more suitable inductive bias for heterogeneous feature interaction. It may also improve parameter efficiency by avoiding exhaustive token-token mixing and concentrating model capacity on a compact set of high-value interaction subspaces.

In this paper, we propose \model, a latent hub-based token mixing architecture for recommendation. The core component of \model is the \model block, which introduces a small set of learnable hubs and organizes efficient information interaction through three stages. The first stage, \emph{hub induction}, allows hubs to selectively aggregate information from heterogeneous tokens. The second stage, \emph{hub interaction}, performs high-order interaction among the induced hubs. The third stage, \emph{token-conditioned readout}, lets each original token query the interacted hubs and retrieve token-specific global interaction signals, thereby preserving field-specific token identity while progressively injecting higher-order latent interaction semantics.

Rather than explicitly modeling a dense $T \times T$ interaction among all feature tokens, each \model block routes information through $H$ latent hubs, where $H \ll T$. This perspective explains why \model can achieve strong performance under smaller parameter budgets: latent hubs provide an explicit inductive bias for organizing sparse and heterogeneous feature interactions. By stacking \model blocks, the same factorized interaction is repeatedly applied to progressively refined token representations, allowing shallow layers to capture low-order cross-field patterns and deeper layers to refine higher-order interaction semantics.

Our contributions are summarized as follows:
\begin{itemize}
    \item We revisit feature interaction in recommendation from the perspective of token mixing. We argue that directly mixing semantically diverse feature tokens can be parameter-inefficient, since useful interactions are often sparse, sample-specific, and group-structured.
    \item We propose \model, a progressive latent hub mixing architecture for parameter-efficient feature interaction. Each \model block follows an induction--interaction--readout paradigm: it induces compact latent hubs from feature tokens, performs interaction in the hub space, and refines each token through token-conditioned readout.
    \item We conduct extensive offline experiments, ablation studies, and online A/B testing in the Kuaishou short-video recruitment business. Offline results show that \model outperforms strong token-mixing baselines with fewer parameters. Online A/B testing further shows a statistically significant 5.48\% improvement in resume submission conversion rate, and \model has been fully deployed in production.
\end{itemize}

\section{Related Work}

\subsection{Feature Interaction in Recommendation}

Feature interaction modeling is a long-standing problem in recommender systems. Classical models such as factorization machines\cite{rendle2010factorization} explicitly model pairwise feature interactions, while neural ranking models such as DCN\cite{wang2017deep}, DCNv2\cite{wang2021dcn}, and AutoInt\cite{song2019autoint} improve interaction modeling through cross networks or attention mechanisms. Recent large-scale recommendation models such as Wukong\cite{zhang2024wukong} further demonstrate the importance of scalable model capacity for industrial ranking.

However, as industrial recommender systems incorporate increasingly diverse feature groups and long behavior sequences, interaction modeling becomes more challenging. The number of possible token-token interactions grows rapidly, while useful interactions are often sparse and context-dependent. Different from existing feature interaction models, \model organizes heterogeneous feature tokens through compact latent hubs, aiming to improve the effectiveness and parameter efficiency of cross-feature interaction.

\subsection{Token-Mixing Architectures for Recommendation}

Transformers have demonstrated strong representational capacity across various domains, including recommendation. Recent studies, such as Hyformer\cite{huang2026hyformer} and OneTrans\cite{zhang2026onetrans}, introduce attention mechanisms into recommendation models and significantly improve their performance. However, the quadratic computational cost of standard self-attention limits its scalability. To improve efficiency, RankMixer\cite{zhu2025rankmixer} adopts the MLP-Mixer architecture and improves both model performance and scalability in industrial ranking. TokenMixer-Large\cite{jiang2026tokenmixer} further introduces token-mixing and reverting operations to improve the scaling capability of large ranking models.

Our work is complementary to this line. Rather than competing primarily on hardware utilization, \model focuses on improving interaction effectiveness and parameter efficiency for heterogeneous recommendation features. It retains a compact latent interaction space while introducing content-adaptive routing through learnable hubs.

\subsection{Latent Query and Cross-Attention Architectures}

Latent architectures introduce a small set of latent variables to summarize large input sets or sequences. Representative examples include Perceiver-style latent vector\cite{jaegle2021perceiver} and query-based designs. Cross-attention provides an asymmetric interface between two representation sets: one set serves as queries and selectively reads information from another set serving as keys and values. This property makes cross-attention suitable for building compact abstraction interfaces, where a small number of latent queries extract task-relevant signals from a larger input set.

A particularly relevant instance is Q-Former\cite{li2023blip}, which bridges a frozen image encoder and a frozen language model through a small set of learnable queries. These queries attend to image features via cross-attention and are then read by the language model, effectively learning a lightweight information bottleneck between two modality-specific encoders. Q-Former demonstrates that a compact set of learnable queries can serve as an effective abstraction interface.

Inspired by the Q-Former framework, OneRec\cite{deng2025onerec} employs learnable queries in its lifecycle path to attend to long user behavior sequences. It inductively summarizes users' lifetime preferences into fixed-size representations, adaptively extracting task-relevant signals from high-cardinality inputs while balancing representation capacity and computational efficiency.

Different from Q-Former and OneRec, which mainly use learnable queries to compress modality features or long behavior sequences, \model uses latent hubs as intermediate interaction centers for general recommendation feature tokens. The goal is not only representation compression, but also structured cross-feature interaction followed by token-conditioned information writing.

\section{Method}

\subsection{Overall Architecture}

\begin{figure*}[t]
  \centering
  \includegraphics[width=\textwidth]{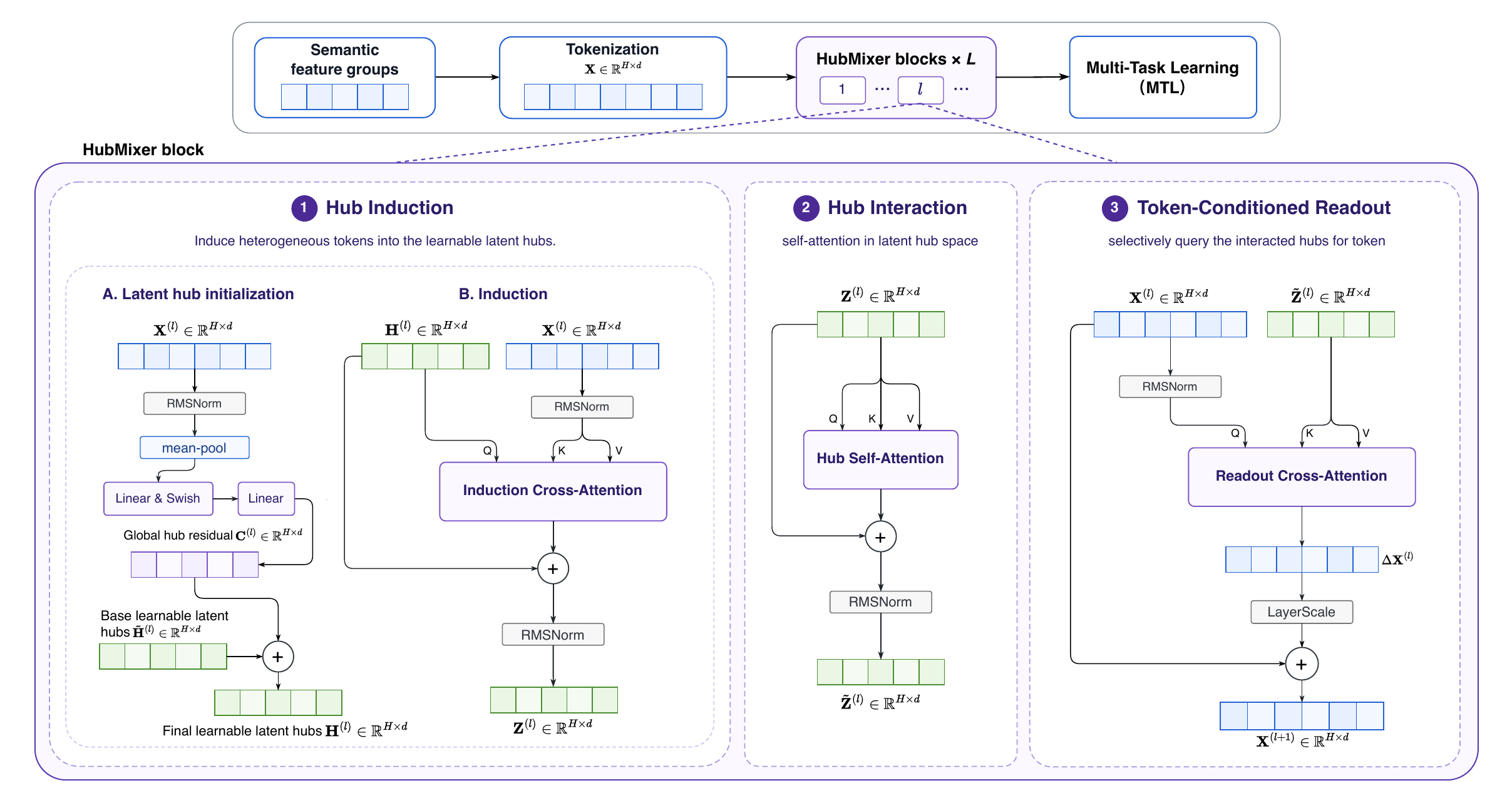}
    \caption{Overall architecture of \model. Heterogeneous feature tokens are first induced into compact learnable latent hubs, then interact in the hub space, and finally selectively query the interacted hubs for token-specific readout.}
    \label{fig:overview}
\end{figure*}

\model is a latent hub-based token mixing architecture for parameter-efficient feature interaction in recommendation system. As illustrated in Figure~\ref{fig:overview}, the architecture follows a four-stage pipeline: tokenization, hub induction, hub interaction, and token-conditioned readout.

\subsection{Tokenization}

Recommendation ranking models usually involve semantically diverse feature groups, such as user profiles, behavior sequences, item features, context signals, and statistical features. Instead of flattening all feature embeddings into an unstructured vector, we organize features according to their semantic groups and transform them into a set of feature tokens:
\begin{equation}
    \mathbf{X} = [\mathbf{x}_1, \mathbf{x}_2, \dots, \mathbf{x}_T] \in \mathbb{R}^{T \times d},
\end{equation}
where $T$ is the number of feature tokens and $d$ is the token dimension. Each token corresponds to a specific semantic feature group or field-level representation. This tokenized representation preserves feature semantics and provides a structured input space for subsequent feature interaction modeling. The goal of \model is to transform $\mathbf{X}$ into refined token representations by progressively modeling interactions among these semantically organized feature tokens.


\model consists of stacked \model blocks. Each block receives token representations $\mathbf{X}^{(l)}$ and outputs refined representations $\mathbf{X}^{(l+1)}$. The \model block first induces compact latent hubs from input tokens, then performs hub-level interaction, and finally writes the interacted information back to tokens through token-conditioned readout. This design keeps the block lightweight while improving parameter efficiency.

Let $\tilde{\mathbf{H}}^{(l)} \in \mathbb{R}^{H \times d}$ denote $l$-th $H$ base learnable latent hubs, where $H \ll T$. These base hubs are shared across instances and optimized end-to-end. They serve as semantic routers that organize heterogeneous feature tokens into compact interaction centers. 
To inject sample-level context into the otherwise static prototypes, we augment the learnable latent hubs with an input-conditioned residual generated by a two-layer MLP:
\begin{equation}
    \mathbf{p}^{(l)}=\frac{1}{T} \sum_{t=1}^{T} \mathbf{x}^{(l)}_t,\qquad
\end{equation}
\begin{equation}
    \mathbf{C}^{(l)}=\phi\!\big(\mathbf{W}^{(l)}_2 \sigma(\mathbf{W}^{(l)}_1  \mathbf{p}^{(l)} +\mathbf{b}^{(l)}_1)+\mathbf{b}^{(l)}_2\big)
\end{equation}
where $\mathbf{C}^{(l)} \in \mathbb{R}^{H \times d}$ is an input-conditioned hub residual generated from the global token summary, ${T}$ denotes the number of tokens, $\sigma(\cdot)$ denotes the Swish activation, $\phi(\cdot)$ reshapes the vector to $\mathbb{R}^{H \times d}$.

Therefore, the final equation of the learnable latent hubs $\mathbf{H}^{(l)}$ is as follows:
\begin{equation}
    \mathbf{H}^{(l)} = \tilde{\mathbf{H}}^{(l)} + \mathbf{C}^{(l)}
\end{equation}

\subsection{Hub Induction}

The first stage summarizes raw heterogeneous tokens into latent hubs. Inspired by Q-Former~\cite{li2023blip}, we use a small set of learnable latent hubs as queries to selectively extract information from input feature tokens. Given normalized token representations $\tilde{\mathbf{X}}^{(l)} = \mathrm{RMSNorm}(\mathbf{X}^{(l)})$, Hub Induction performs cross-attention from tokens to hubs, aiming to induct token representations into the learnable latent hubs $\mathbf{H}^{(l)}$:

\begin{equation}
    \mathbf{Z}^{(l)} = \mathrm{RMSNorm}(\mathbf{H}^{(l)} + \mathrm{CrossAttn}(\mathbf{Q}=\mathbf{H}^{(l)}, \mathbf{K}=\tilde{\mathbf{X}}^{(l)}, \mathbf{V}=\tilde{\mathbf{X}}^{(l)})),
\end{equation}

where $\mathbf{Z}^{(l)} \in \mathbb{R}^{H \times d}$ is the induced hub representation.

Intuitively, each hub learns to attend to the mixture of feature tokens. Different hubs can specialize in different semantic aspects, such as user interest, item semantics, context signals, and user-item interaction. Because the attention weights depend on input content, Hub Induction provides sample-specific routing from heterogeneous feature tokens to latent interaction centers.

\subsection{Hub Interaction}

After induction, \model performs interaction in the compact hub space. We apply self-attention to the induced hubs to model hub-level dependencies: 
\begin{equation}
    \tilde{\mathbf{Z}}^{(l)} =
    \mathrm{RMSNorm}(
        \mathbf{Z}^{(l)} +
        \mathrm{SelfAttn}(
            \mathbf{Q}=\mathbf{Z}^{(l)},
            \mathbf{K}=\mathbf{Z}^{(l)},
            \mathbf{V}=\mathbf{Z}^{(l)}
        )
    ).
\end{equation}
Since the number of hubs is small, hub-level interaction is computationally efficient.


This stage is crucial because it models high-order feature interactions after token information has been selectively aggregated into a small number of latent hubs. In recommendation ranking, raw feature tokens come from different semantic groups and many token pairs are weakly related for a given request. Directly modeling all token-token dependencies therefore spends capacity on a large number of less useful interactions. Hub Interaction instead operates on induced hubs, where each hub already summarizes a content-dependent subset of relevant feature signals. Modeling dependencies among these hubs provides a structured inductive bias for capturing compact and high-value cross-feature interactions.




\subsection{Token-Conditioned Readout}

After hub-space interaction, the interacted hubs encode compact global cross-feature semantics. A straightforward way to use these hubs is to pool them into a single global vector and feed it to the prediction module. However, such a design may collapse field-specific token identities, which are important in recommendation ranking where different feature tokens play different roles for different objectives. Therefore, \model projects the interacted hub semantics back to the token space through token-conditioned readout.

Specifically, each token uses its normalized representation as a query to selectively read from the interacted hubs:
\begin{equation}
\Delta \mathbf{X}^{(l)}
=
\mathrm{CrossAttn}
\left(
\mathbf{Q}=\tilde{\mathbf{X}}^{(l)},
\mathbf{K}=\tilde{\mathbf{Z}}^{(l)},
\mathbf{V}=\tilde{\mathbf{Z}}^{(l)}
\right),
\end{equation}
where $\Delta \mathbf{X}^{(l)} \in \mathbb{R}^{T \times d}$ is the token-specific readout signal. Different from global pooling or broadcasting, this operation allows each token to retrieve a customized mixture of latent hub semantics according to its own representation.

The readout signal is then injected into the original token stream through a LayerScale-style residual connection:
\begin{equation}
\mathbf{X}^{(l+1)}
=
\mathbf{X}^{(l)}
+
\boldsymbol{\gamma}^{(l)} \odot \Delta \mathbf{X}^{(l)},
\end{equation}
where $\boldsymbol{\gamma}^{(l)} \in \mathbb{R}^{d}$ is a learnable LayerScale parameter and $\odot$ denotes element-wise multiplication with broadcasting over the token dimension. This residual formulation preserves the original field-specific token representations while allowing the model to adaptively inject interaction-aware hub information. In practice, LayerScale provides a stable mechanism to control the magnitude of the readout update, which is especially useful when stacking multiple \model blocks.

Consequently, the output of the $l$-th block is not a pooled hub representation, but a refined token set $\mathbf{X}^{(l+1)}$. This design keeps token-level feature identities available for downstream multi-task prediction, while progressively enriching each token with global cross-feature interaction signals from the latent hub space.

\subsection{Multi-Task Prediction and Optimization}

We train \model end-to-end with a multi-task learning (MTL) model. Let $\mathcal{K}$ denote the set of ranking tasks. For each ranking task $k$, we denote the task label as $y_{k}$, the prediction as $\hat{y}_{k}$.

Given the output tokens from the final \model block, the enhanced token representations are fed into the MTL module for simultaneous optimization of multiple ranking objectives. These tokens are first aggregated into an instance-level representation:
\begin{equation}
    \mathbf{r} = \mathrm{Concat}(\mathbf{X}^{(L)}).
\end{equation}
where $\mathrm{Concat}(\cdot)$ is implemented as the concatenation of all token representations. Each task head then produces its own prediction:
\begin{equation}
    \hat{y}_k = \mathrm{Sigmoid}_k(\mathrm{MTL}_k(\mathbf{r})),
\end{equation}
where $k$ indexes the ranking task.

The binary classification cross-entropy loss function is computed as the loss for the ranking task $k$:
\begin{equation}
    \ell_k(y_{k}, \hat{y}_{k}) =
    - y_{k} \log \hat{y}_{k}
    - (1-y_{k}) \log (1-\hat{y}_{k}).
\end{equation}
where $\ell_k(\cdot)$ is the task-specific loss function. 

The overall objective is a weighted sum over all tasks:
\begin{equation}
    \mathcal{L} =
    \sum_{k=1}^{|\mathcal{K}|} \lambda_{k} \cdot \ell_k(y_{k}, \hat{y}_{k}),
\end{equation}
where $\lambda_k$ controls the relative importance of each task. \model handles cross-feature interaction, and the MTL model handles task-specific signal extraction. 

\section{Experiments}

\subsection{Experimental Setup}

\paragraph{Experiment Overview.}
We evaluate \model on a large-scale industrial dataset collected from the Kuaishou short-video recruitment business. This business is a revenue-generating blue-collar recruitment service built upon Kuaishou's content distribution scenario, where job-related short videos and recruitment leads are distributed to potentially interested users, covering around 40 million daily active users. The ranking model is optimized for lead generation, aiming to improve both the quantity of resumes submitted by users and downstream recruitment conversion.

\paragraph{Data Collection and Objectives Optimization}
 We collected one-month behavioral data on potentially interested users from Kuaishou’s large-scale logs and created a dataset of over 1 billion samples. The dataset contains user profiles, user behavior sequence features, job features, short-video features, contextual features, and statistical features. Following the production setting, we formulate the task as a multi-task ranking problem with four prediction targets: \texttt{plc\_click}, \texttt{effective\_view}, \texttt{interact}, and \texttt{resume\_submit}. These objectives correspond to different stages of the recruitment, from initial click and effective consumption to interaction and final resume submission. 

\paragraph{Baselines.}
We compare \model with representative industrial feature interaction and token-mixing ranking models, including DCN, DCNv2, AutoInt, Wukong, RankMixer, and TokenMixer. These baselines cover explicit feature crossing models, attention-based feature interaction models, interaction-oriented industrial ranking modules, and recent token-mixing ranking backbones. All models are trained on the same feature set and optimized under the same multi-task setting for fair comparison.

\paragraph{Metrics.}
For offline evaluation, we use AUC to assess the model's performance; a higher AUC value reflects the model's ranking ability. Since the four tasks correspond to different stages, we report per-task AUC rather than collapsing all objectives into a single score. For online experiments, our primary online metric is the resume submission conversion rate. However, since users' willingness to submit resumes is influenced by factors such as click, effective consumption, and interaction, we combine the four task predictions into a comprehensive ranking score to model users' resume-submission intent from a holistic perspective.

\paragraph{Implementation Details.}
Unless otherwise specified, all models use the same feature preprocessing pipeline, embedding dimension, optimizer, batch size, training schedule, and multi-task loss weights. 
For all Mixer models, the number of tokens is set to 32, and the number of Mixer block layers is set to 2. Specifically, our \model's default number of hubs is set to $H=16$.

\subsection{Performance Comparison}

\begin{table*}[t]
\centering
\caption{Performance comparison on the Kuaishou short-video recruitment dataset. We report AUC and parameter count on four multi-task ranking objectives.}
\label{tab:main_results}
\begin{tabular}{lcccccc}
\toprule
Model & \texttt{plc\_click} & \texttt{effective\_view} & \texttt{interact} & \texttt{resume\_submit} & Avg. AUC & \#Params  \\
\midrule
DCN     &       0.8196 & 0.8453 & 0.8358 & 0.7818 & 0.8206 & 155.3M \\
DCNv2   &       0.8194 & 0.8468 & 0.8361 & 0.7825 & 0.8212 & 157.5M \\
AutoInt &       0.8194 & 0.8459 & 0.8363 & 0.7823 & 0.8210 & 162.6M \\
Wukong &       0.8183 & 0.8461 & 0.8365 & 0.7822 & 0.8208 & 154.6M \\
RankMixer &     0.8221 & \underline{0.8495} & 0.8389 & 0.7845 & 0.8238 & 155.1M \\
TokenMixer &    \underline{0.8227} & 0.8491 & \underline{0.8394} & \underline{0.7852} & \underline{0.8241} & 156.9M \\
\midrule
\model(ours) &  \textbf{0.8253} & \textbf{0.8507} & \textbf{0.8401} & \textbf{0.7864} & \textbf{0.8256} & 142.4M \\
\bottomrule
\end{tabular}
\end{table*}

Table~\ref{tab:main_results} compares \model with representative industrial feature interaction and token-mixing models, including DCN, DCNv2, AutoInt, Wukong, RankMixer, and TokenMixer. These models cover widely used feature crossing architectures as well as recent token-mixing backbones for recommendation ranking. We make the following observations.

First, traditional feature crossing models, such as DCN, DCNv2, AutoInt, and Wukong, generally underperform recent token-mixing ranking models such as RankMixer, TokenMixer. This suggests that directly increasing the order or form of feature crossing is not sufficient for modern industrial ranking scenarios, where feature tokens are semantically diverse and useful interactions are sparse and sample-dependent.

Second, \model achieves the best AUC on all four ranking objectives. The improvements are consistent across both early-stage engagement targets and deeper conversion-oriented targets, indicating that \model learns interaction patterns that are useful for the whole recruitment rather than for a single objective only.


Third, \model uses fewer parameters than RankMixer and TokenMixer while achieving better AUC. This result highlights the parameter efficiency of the proposed induction--interaction--readout paradigm, meaning that routing feature interactions through compact latent hubs can use model capacity more effectively than directly mixing all feature tokens in a flat token space.

\subsection{Ablation Study}

\paragraph{Module Ablation.}
We first evaluate the contribution of key \model components. We compare the full model with two variants: (1) removing hub interaction, where induced hubs are directly used for readout without hub-space self-attention; and (2) replacing token-conditioned readout with pooled hub injection, where the interacted hubs are pooled into a global vector and added to all tokens. The second variant tests whether token-specific selective readout is necessary, or whether a shared global summary is sufficient.

\begin{table*}[t]
\centering
\caption{Module ablation on the Kuaishou short-video recruitment dataset.}
\label{tab:ablation_components}
\begin{tabular}{lccccc}
\toprule
Variant & \texttt{plc\_click} & \texttt{effective\_view} & \texttt{interact} & \texttt{resume\_submit} & Avg. AUC \\
\midrule
Full \model & 0.8253 & 0.8507 & 0.8401 & 0.7864 & 0.8256 \\
w/o Hub Interaction & 0.8221 & 0.8480 & 0.8383 & 0.7844 & 0.8232 \\
Pooling-to-Token Readout & 0.8242 & 0.8499 & 0.8390 & 0.7857 & 0.8247 \\
\bottomrule
\end{tabular}
\end{table*}

Table~\ref{tab:ablation_components} reports the module ablation results. The full \model consistently achieves the best AUC across all four tasks, validating the effectiveness of the complete induction--interaction--readout design. Removing hub interaction leads to clear performance degradation on every objective, with the average AUC dropping from 0.8256 to 0.8232. This indicates that simply inducing latent hubs is not sufficient; high-order interaction among hubs is necessary to transform the induced summaries into useful cross-feature interaction signals.

Replacing token-conditioned readout with pooled hub injection also hurts performance, although the degradation is smaller than removing hub interaction. This result suggests that global hub information is useful, but uniformly adding a pooled hub representation to all tokens is less effective than letting each token selectively retrieve relevant information from the interacted hubs. In other words, token-conditioned readout preserves token-specific semantics while injecting interaction context, which is important for multi-task recruitment ranking where different feature tokens may contribute differently to click, effective view, interaction, and resume submission objectives.

\paragraph*{Effect of Hub Number.}
We further study the effect of the number of latent hubs. The hub number controls the capacity of the latent interaction space: too few hubs may impose an excessive information bottleneck, while too many hubs increase parameters and computation with potentially diminishing returns. We set $H=16$ as the default configuration and compare it with $H=4$, $H=8$, and $H=32$.

\begin{table*}[t]
\centering
\caption{Effect of the number of latent hubs. The default setting is $H=16$.}
\label{tab:hub_number}
\begin{tabular}{lcccccc}
\toprule
$H$ & \texttt{plc\_click} & \texttt{effective\_view} & \texttt{interact} & \texttt{resume\_submit} & Avg. AUC & Params \\
\midrule
4           & 0.8231 & 0.8486 & 0.8382 & 0.7852 & 0.8238 & 136.1M \\
8           & 0.8244 & 0.8498 & 0.8394 & 0.7857 & 0.8248 & 138.3M \\
16(default) & 0.8253 & 0.8507 & 0.8401 & 0.7864 & 0.8256 & 142.4M \\
32          & 0.8255 & 0.8511 & 0.8402 & 0.7865 & 0.8258 & 150.8M \\
\bottomrule
\end{tabular}
\end{table*}

Table~\ref{tab:hub_number} shows that increasing the number of hubs from 4 to 16 consistently improves all four task AUCs, indicating that a larger latent hub space provides stronger capacity for organizing cross-feature interactions. However, further increasing the hub number from 16 to 32 yields only marginal gains: the average AUC improves from 0.8256 to 0.8258, while the parameter count increases from 142.4M to 150.8M. This suggests that the information gain begins to saturate beyond 16 hubs. Therefore, we choose $H=16$ as the default setting, which provides a better accuracy-parameter trade-off for the recruitment ranking task.

\subsection{Online A/B Test}

To further validate the effectiveness of \model in real production environments, we deployed \model in the Kuaishou content-based recruitment short-video distribution system. The online A/B test lasted for 7 days and covered 7.2\% of the production traffic. The primary online metric is the resume submission conversion rate, which directly reflects the downstream lead-generation objective of the recruitment business.

\model improves the overall resume submission conversion rate by 5.48\% over the base model, and the improvement is statistically significant. This online gain demonstrates that the proposed induction--interaction--readout architecture not only improves offline AUC, but also translates into measurable business impact in large-scale short-video recruitment distribution. Notably, after the A/B test, \model has been fully deployed in the Kuaishou short-video recruitment business.

\subsection{Analysis of Token Representation Enhancement}

To further understand why \model outperforms strong token-mixing baselines, we conduct a detailed representation analysis against TokenMixer, the strongest competing baseline in our performance comparison. Since TokenMixer also operates on tokenized feature representations, this comparison provides a controlled view of how different token interaction mechanisms modify and enhance feature tokens.

We analyze token representations from two complementary perspectives: token-wise representation updates and linear probing performance. The former measures how much each mixer block changes the direction of each token representation, while the latter examines whether the enhanced representations contain more accessible task-relevant information.

For the $l$-th block, let $\mathbf{x}_{n,i}^{(l)}$ and $\mathbf{x}_{n,i}^{(l+1)}$ denote the input and output representations of the $i$-th token for instance $n$, respectively. We measure the token-wise directional update by cosine distance:
\begin{equation}
    D_i^{(l)}
    =
    \mathbb{E}_{n}
    \left[
    1 -
    \cos
    \left(
    \mathbf{x}_{n,i}^{(l)},
    \mathbf{x}_{n,i}^{(l+1)}
    \right)
    \right].
\end{equation}
A larger value of $D_i^{(l)}$ means that the output token becomes less similar to its input representation after block-level interaction, indicating a larger amount of newly injected interaction information.

\begin{figure}[t]
    \centering
    \includegraphics[width=0.47\textwidth]{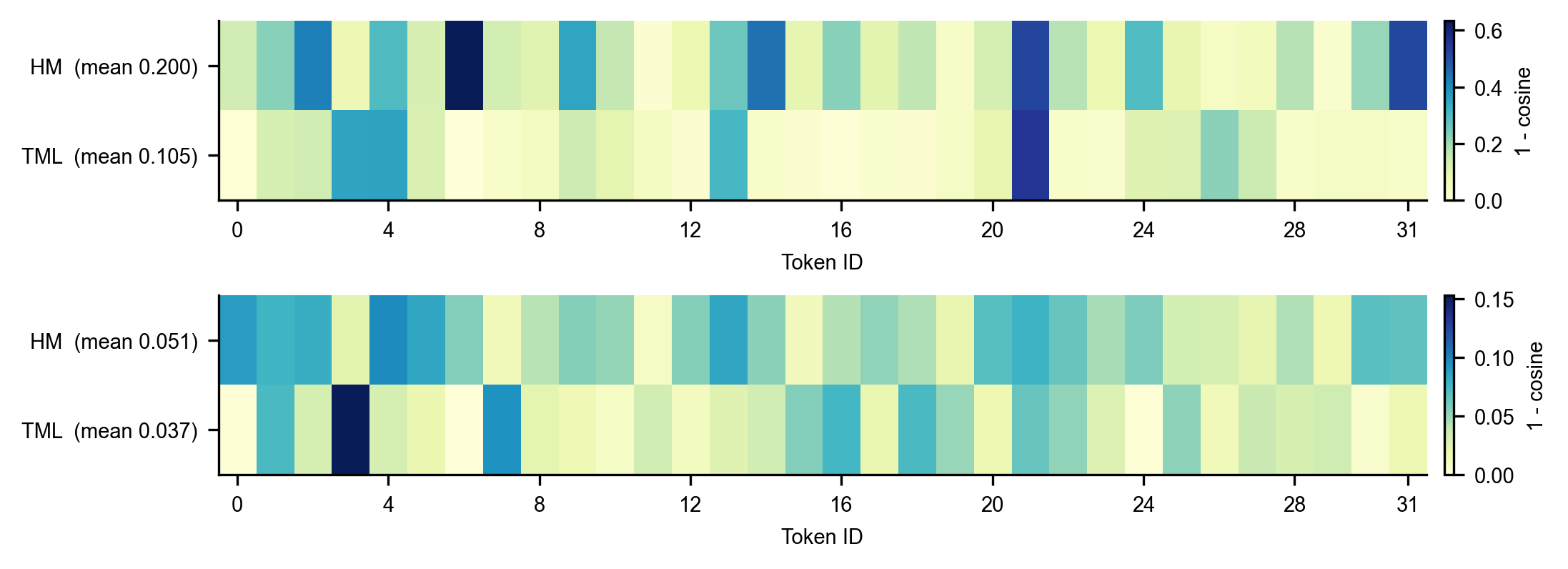}
    \caption{Token-wise input-output cosine distance of \model and TokenMixer across mixer layers. A larger value indicates a stronger directional update from block input to block output.}
    \label{fig:representation_update}
\end{figure}

As shown in Figure~\ref{fig:representation_update}, \model presents consistently larger input-output dissimilarity than TokenMixer. In the first mixer layer, the average cosine distance of \model is 0.200, nearly twice that of TokenMixer (0.105). In the second layer, \model still maintains a larger average cosine distance (0.051 vs. 0.037). These results indicate that, after block-level interaction, \model changes the direction of token representations more substantially than TokenMixer.

This observation is consistent with the design motivation of \model. TokenMixer directly mixes tokens in the original token space, which may lead to relatively mild token updates. In contrast, \model first induces compact latent hubs from the input tokens, performs high-order interaction in the hub space, and then writes the interacted information back to each token through token-conditioned readout. Therefore, the larger input-output dissimilarity suggests that \model can inject richer interaction information into the corresponding token representations. Meanwhile, the heatmap also shows that the updates are not uniformly distributed over all tokens, implying that the injected information is token-dependent rather than a simple global perturbation.

However, larger representation updates alone do not necessarily imply better representation quality, since they may also come from unnecessary perturbation. Therefore, we further conduct linear probing to evaluate whether the enhanced token representations are more predictive for downstream ranking tasks. Specifically, after training the backbone model, we freeze all backbone parameters and train lightweight linear classifiers on top of each token representation. We also evaluate an instance-level probe on the concatenated representation $\mathrm{Concat}(\mathbf{X}^{(L)})$ to measure the overall representation quality.

\begin{figure}[t]
    \centering
    \includegraphics[width=0.47\textwidth]{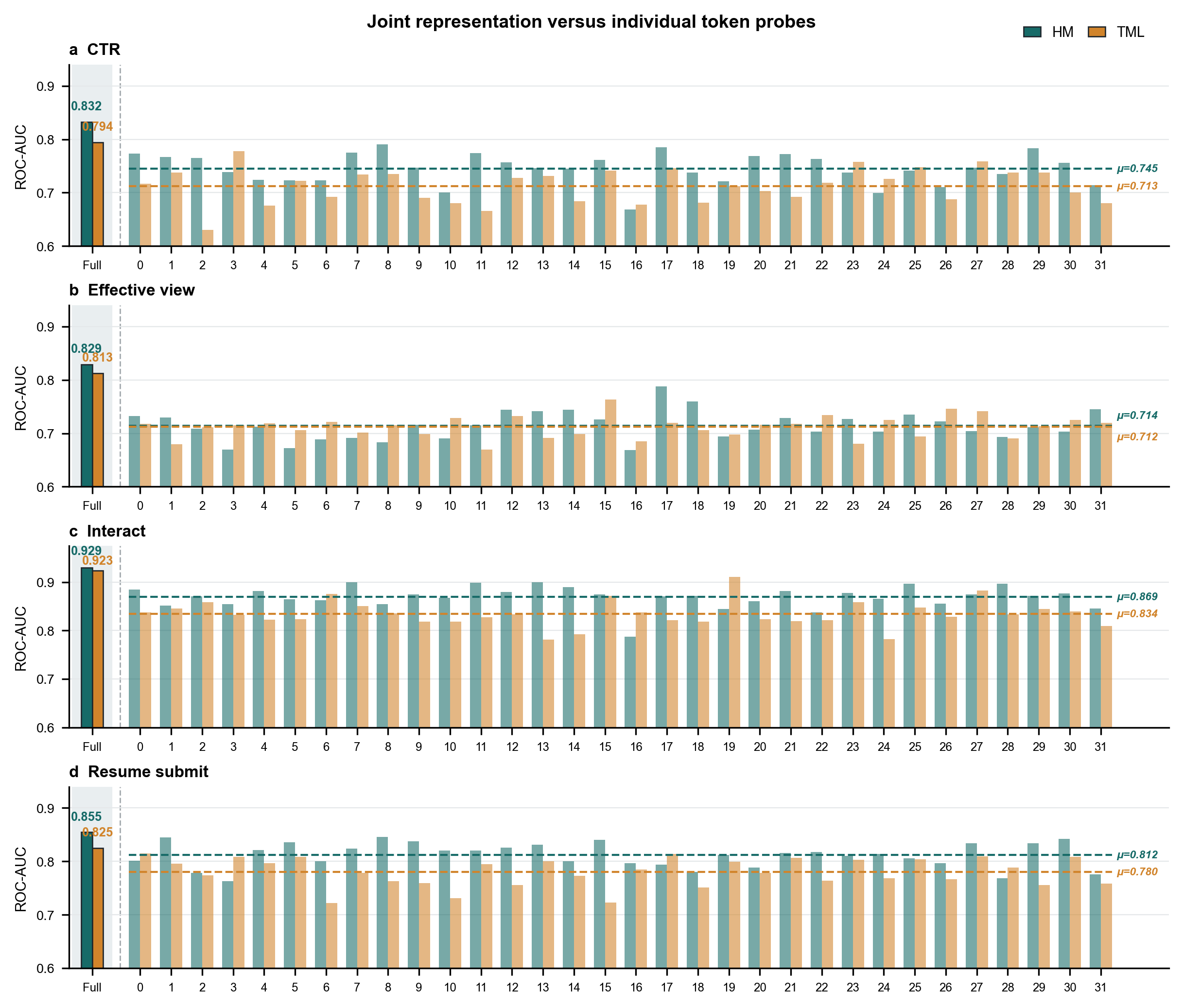}
    \caption{Linear probing comparison between \model and TokenMixer. The backbone is frozen, and lightweight linear classifiers are trained on token-level or concatenated instance-level representations.}
    \label{fig:linear_probe}
\end{figure}

As shown in Figure~\ref{fig:linear_probe}, \model achieves higher probing AUC than TokenMixer on most feature tokens and also obtains better probing performance on the aggregated representation. Since the probing classifiers are linear and the backbone is frozen, the improvement indicates that \model makes task-relevant signals more accessible. This result complements the cosine-distance analysis: compared with TokenMixer, \model not only modifies token representations more strongly, but also produces representations that are more predictive for multi-task recruitment ranking.

Together, these results suggest that the performance gain of \model over TokenMixer is not simply explained by stronger representation perturbation. Instead, the induction--interaction--readout paradigm enhances token representations by injecting task-relevant cross-feature interaction information, which helps explain the superior AUC of \model in the main performance comparison.

\section{Discussion}

\subsection{Latent Hubs as an Inductive Bias}

The proposed latent hubs can be viewed as an architectural inductive bias for organizing recommendation features before explicit interaction. In industrial ranking systems, feature tokens often come from different semantic sources, such as user profiles, behavior sequences, item attributes, context signals, and statistical features. Treating these tokens as a flat sequence leaves the model to discover useful interaction structures entirely from data. \model instead introduces a small set of learnable interaction centers, encouraging the model to first summarize related signals into compact latent semantics and then perform interaction among these semantics. This design does not assume a fixed feature hierarchy; rather, it lets the hubs learn soft, content-dependent organization patterns during end-to-end training.

\subsection{Readout as Selective Information Writing}

The token-conditioned readout stage is better understood as selective information writing rather than global feature broadcasting. The interacted hubs store compact cross-feature semantics, while the original tokens preserve field-specific information required by downstream prediction heads. Token-conditioned readout connects these two spaces: each token decides which hub information to absorb according to its own query, producing enhanced token representations for final prediction. This design is particularly suitable for multi-task ranking, because different objectives may rely on different aspects of the same feature set. For example, an engagement-oriented target and a conversion-oriented target may benefit from different combinations of user intent, item semantics, and contextual signals. Selective readout injects these cross-feature interaction signals back into token representations, allowing the downstream multi-task heads to extract useful evidence from the enhanced token set without collapsing all tokens into a single global summary.

\subsection{Practical Deployment Considerations}

From a deployment perspective, \model provides several practical knobs for industrial ranking systems. The hub number controls the size of the latent interaction space and can be adjusted according to the available parameter and latency budget. The number of \model blocks controls the depth of progressive interaction refinement. In addition, both hub induction and token-conditioned readout are formulated as cross-attention between feature tokens and a small set of hubs, whose cost scales with $TH$ rather than $T^2$. This structure is amenable to engineering optimization such as batched cross-attention, kernel fusion, and optimized small-matrix computation. These considerations suggest that the induction--interaction--readout design is not only a modeling principle, but also a flexible framework that can be adapted to different production constraints.

\subsection{Future Directions}

\model opens several directions for future exploration. First, the current design uses a shared set of learnable hubs across all instances through a lightweight input-conditioned residual. A stronger extension is to generate the full hub set dynamically from request-level context, user intent, or scenario-specific signals. Such adaptive hubs may further improve the flexibility of latent routing, especially in production systems where traffic distributions and user intents are highly diverse.

Second, hub specialization can be further encouraged. Although learnable hubs can naturally develop different semantic preferences during end-to-end training, explicit diversity regularization or orthogonality constraints may help different hubs cover complementary feature groups and interaction patterns. This direction is particularly relevant when stacking more \model blocks, where maintaining diverse hub semantics may improve progressive interaction refinement.

Third, \model can be combined with more advanced multi-task learning designs. In this work, \model serves as a shared feature interaction backbone before task-specific prediction heads. Future studies may explore task-aware readout, task-conditioned hubs, or objective-specific routing, so that different ranking objectives can selectively access different interaction semantics while still sharing the same latent hub space.

Fourth, the hub interaction operator itself can be made more flexible. The current implementation adopts self-attention over latent hubs to model high-order hub-level dependencies. Since the number of hubs is much smaller than the number of feature tokens, this cost is already controlled; nevertheless, for scenarios with stricter FLOPs or latency budgets, the hub interaction module could be replaced by more lightweight operators, such as MLP-based hub mixing or sparse expert-style hub mixing. These alternatives may further reduce computational complexity while preserving the overall induction--interaction--readout framework.

Finally, system-level optimization is important for large-scale deployment. Since both hub induction and token-conditioned readout are formulated as cross-attention between feature tokens and a small set of latent hubs, their computation scales with $TH$ rather than $T^2$ when $H \ll T$. This structure provides practical opportunities for batching-friendly cross-attention implementation, kernel fusion, and optimized small-matrix computation. These engineering directions can further improve serving efficiency while preserving the selective induction--interaction--readout mechanism.

\section{Conclusion}

In this work, we propose \model, a progressive latent hub mixing architecture for parameter-efficient feature interaction in recommendation. \model introduces a small set of learnable latent hubs through a lightweight input-conditioned residual and organizes cross-feature interaction through an induction--interaction--readout paradigm. The hubs first selectively aggregate semantically diverse feature tokens, then perform high-order interaction in the compact hub space, and finally each token retrieves a customized interaction signal through token-conditioned readout. This design forms a content-adaptive interaction mechanism that concentrates model capacity on structured, high-value cross-feature patterns rather than diffusing it over all token pairs.

We further show that the readout output serves directly as the block output and is fed into multi-task prediction heads. This design keeps cross-feature interaction and multi-task signal extraction cleanly separated: \model focuses on producing enhanced token representations, while the downstream prediction heads extract task-specific signals for final objectives. When stacked across depth, the progressive refinement of token representations enables the model to capture increasingly abstract cross-feature interactions. Representation analysis suggests that \model injects stronger and more task-relevant interaction information into token representations than TokenMixer. Input-output representation analysis further confirms that \model better preserves field-specific token identity while injecting interaction context.

Extensive offline experiments demonstrate that \model achieves better ranking performance than strong token-mixing baselines with fewer parameters. Ablation studies validate the independent contributions of hub interaction, and token-conditioned readout. Online A/B testing further confirms the production value of \model, yielding a statistically significant 5.48\% improvement in resume submission conversion rate, after which \model has been fully deployed in the Kuaishou short-video recruitment business. These results suggest that latent hub mixing provides an effective and extensible paradigm for parameter-efficient feature interaction in industrial recommendation ranking.

\bibliographystyle{ACM-Reference-Format}
\bibliography{main}

\end{document}